\documentclass[sigconf,nonacm]{acmart}

\usepackage{booktabs}
\usepackage{algorithmic}
\usepackage{graphicx}
\usepackage[unicode]{hyperref}
\usepackage{textcomp}
\usepackage{xcolor}
\usepackage[linesnumbered,ruled,vlined]{algorithm2e}

\begin{document}

\title{Efficient global register allocation}

\author{Ian Rogers}
\affiliation{\institution{Google}
  \streetaddress{1600 Amphitheater Parkway}
  \city{Mountain View}
  \state{California}
  \country{USA}
}
\email{irogers@google.com}

\begin{abstract}

In a compiler, an essential component is the register allocator. Two
main algorithms have dominated implementations, graph coloring and
linear scan, differing in how live values are modeled. Graph coloring
uses an edge in an `interference graph' to show that two values cannot
reside in the same register. Linear scan numbers all values, creates
intervals between definition and uses, and then intervals that do not
overlap may be allocated to the same register. For both algorithms the
liveness models are computed at considerable runtime and memory
cost. Furthermore, these algorithms do little to improve code quality,
where the target architecture and register coalescing are important
concerns.

We describe a new register allocation algorithm with lightweight
implementation characteristics. The algorithm introduces a
`future-active' set for values that will reside in a register later in
the allocation. Registers are allocated and freed in the manner of
linear scan, although other ordering heuristics could improve code
quality or lower runtime cost. An advantageous property of the
approach is an ability to make these trade-offs. A key result is the
`future-active' set can remove any liveness model for over 90\%
of instructions and 80\% of methods. The major contribution is the
allocation algorithm that, for example, solves an inability of the
similarly motivated Treescan register allocator \cite{Colombet11} to
look ahead of the instruction being allocated - allowing an
unconstrained allocation order, and an ability to better handle fixed
registers and loop carried values. The approach also is not reliant on
properties of SSA form, similar to the original linear scan work. An
analysis is presented in a production compiler for Java code compiled
through SSA form to Android dex files.

\end{abstract}

\maketitle


\section{Introduction}

The problem at the heart of register allocation is how to allocate
instructions (producing values) to registers so that a register is not
in use, holding the result of two `live' instructions, at the same
time. An approach to modeling this problem is with an interference
graph, where instructions are vertices and edges exist between
vertices live at the same time. This model allows register allocation
to be solved through graph coloring
\cite{Chaitin:1981:RAV:2245737.2245881}, where each color is a
distinct register. An alternate approach is to serialize and
incrementally number instructions, intervals are then formed from the
definition to the last use of an instruction. If two intervals have an empty
intersection then they may be allocated to the same register
\cite{Poletto:1997:TSF:258915.258926, Poletto:1999:LSR:330249.330250}.

Linear scan register allocation has been refined to allow for liveness
holes, and to vary the order the intervals are processed
\cite{Sarkar:2007:ELS:1759937.1759950, Traub:1998:QSL:277650.277714,
  olesen2011register}. Interval based register allocators are popular
due to their performance and for being easy to tweak using
heuristics.

Modeling intervals comes with clear memory costs. Typically an
interval is associated with one or more instructions, and the interval
itself is a collection of pairs of beginning and end integers. As an
interval may be needed for every instruction, in Static Single
Assignment (SSA) form, the interval's memory requirement is often
similar to that of the instruction representation. Phrased another
way, modeling intervals can more than double the compiler's memory
usage.  Graph coloring similarly impacts memory and in his seminal
paper Chaitin concludes with ``a fair amount of virtual storage is
needed to hold the program IL and interference graph,''
\cite{Chaitin:1981:RAV:2245737.2245881}.

As a runtime cost, interval construction is often a significant
portion of register allocation time. Poletto and Sarkar's early linear
scan work shows ``allocation setup'', described as, ``the construction
of live intervals,'' as being the largest portion of time spent for
the register allocation of ``dynamic code kernels'' (Fig. 3 in
\cite{Poletto:1999:LSR:330249.330250}). The overhead of interval
construction is used to motivate a fast interval construction that
unfortunately lowered code quality.

\begin{figure}[bp]
  \vspace{-0.5cm}
  \includegraphics{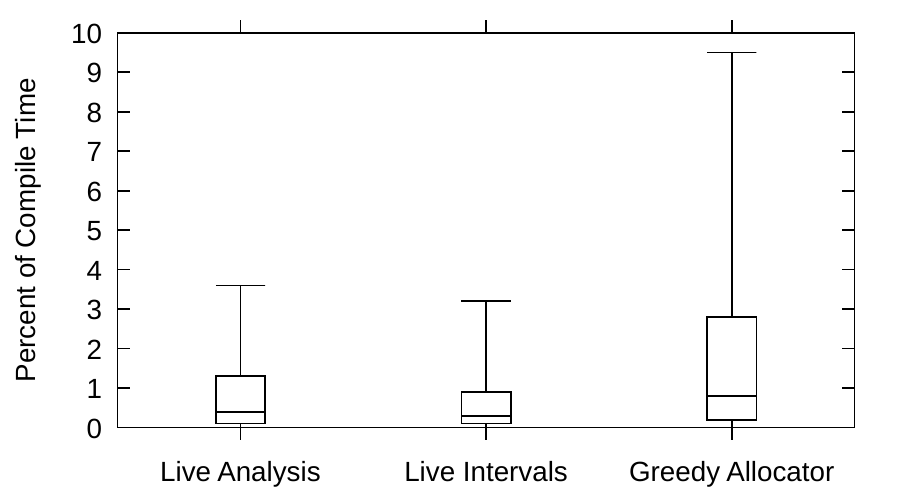}
  \vspace{-0.5cm}
  \caption{Contribution to LLVM compile time of register allocator
    phases. Compilation times were measured using on LLVM 7.0 using
    its time reporting option. Each file was compiled 30 times on an
    Intel Xeon E5-2690 at 2.9GHz with 64GB of RAM.}
  \label{fig:llvm70}
\end{figure}

Fig.~\ref{fig:llvm70} shows a repeat of Poletto and Sarkar's analysis
but for LLVM \cite{Lattner:2004:LCF:977395.977673} compiling itself at
compilation level `-O2'. The box plots show the minimum, 1 percentile,
median, 99 percentile and maximum compile time percentage of each
phase compiling a file from LLVM, where the register allocator is
LLVM's greedy allocator \cite{olesen2011register}. Unlike Poletto and
Sarkar's early work \cite{Poletto:1999:LSR:330249.330250}, interval
construction is not the slowest of the 3 phases. However, removing the
phase would save 0.2\% of compile time or 20\% of one of the major
portions of register allocation time. Just as when Poletto and Sarkar
introduced linear scan, interval construction costs have been
attempted to be avoided in a number of register allocators. We will
review these allocators in section~\ref{sec:related}.

After memory and runtime complexity, the final cost we sought to
eliminate was the implementation complexity. Better memory and runtime
performance were a concern, not because lower code quality was
acceptable but the opposite, we wanted to focus the implementation
effort on best mapping to the target architecture. In common with
LLVM's greedy allocator \cite{olesen2011register},
existing compilers for the target architecture have a register
assignment phase (also known as a rewrite phase) to modify impossible
register allocations to ones that fit the machine's constraints
\cite{bornstein2008dalvik, d8}. We wished to eliminate this rewriting,
as even in simple cases a third of instructions in the final code
could be introduced by it. However, by allocating directly into the
target registers we may need to restart the register allocator, for
example, if register usage had become fragmented and no allowable
register was available. The presented approach analyzes the register
allocation at the point no allocation is available, selects the best
candidate instruction that can have its live range split, inserts
moves and then restarts the register allocation\footnote{Described in
  detail in section~\ref{sec:alloc_failure}.}. The introduced move
instructions become candidates for the existing register coalescing
algorithms. Performing a similar analysis and transformation with
intervals is a challenge. Firstly, intervals and not instructions are
allocated and so determining cost requires going through an
abstraction and determining what intervals and instructions are being
modeled as live at the failure point. Secondly, the liveness model
needs updating following the transformation, something that may be
handled by additional state or by just recomputing the model. It could
be that existing register coalescing performed is impacted by live
range splitting, something the presented approach handles elegantly by
not having state but that requires existing approaches to either
ignore, recompute or model via additional complex state.

Reduced complexity and an ability to spot optimizations are key
advantages to the presented algorithm.  We will discuss the target
machine challenges that motivated them in section~\ref{sec:background}
as well as the compiler framework. These advantages would also be
apparent were the register allocation algorithm used for other
architectures.

Section~\ref{sec:interval-free-ra} describes the major contribution of
the paper, a register allocator algorithm that performs on-the-fly
construction of `future-active' sets and the `live-at-the-same-time'
operations that allow these to function as intervals. For an intuition
as to why this improves performance, the `future-active' sets are only
used when registers are not allocated to the active-set and need to be
accessed later during the register allocation. The code generation
behavior is identical to a regular interval based linear scan register
allocator, and the runtime performance differs only in how liveness is
modeled. Solutions to numerous production challenges, such as exception
handling, are described.

Section~\ref{sec:analysis} analyzes the runtime performance in the
production setting of compiling the entirety of the Android operating
system's Java code. Inspecting per-method or per-register-allocation
metrics is preferred to a shoot-out of one register allocation
algorithm against another. This is because over 80\% of methods are
compiled with no use of the `future-active' sets, for which a
traditional linear scan algorithm may have spent 20\% of its time
computing intervals (from Fig.~\ref{fig:llvm70}). That is the
shoot-out would just highlight the runtime win of not having a
liveness modeling phase.  As such, it is more informative to analyze
what occurs for the new algorithm in the 20\% of methods that need the
`future-active' sets and to see if they do or do not motivate the use
of intervals compared to the new approach. Similarly, comparisons are
not made against algorithms such as graph coloring as this would not
address implementation complexity and such comparisons are already
readily available in existing literature
\cite{Poletto:1999:LSR:330249.330250, Mossenbock02,
  Sarkar:2007:ELS:1759937.1759950}.

The final sections of the paper consider improvements that can further
reduce cost and/or that can increase code quality. A comparison of the
most related work is performed and finally conclusions are drawn.

\section{Background}
\label{sec:background}

In the compiler we model a register allocator as taking an instruction
and mapping it to a target machine register. We present the target machine
in section~\ref{sec:dex} and the compiler in
section~\ref{sec:appreduce}.

\subsection{Dex files}
\label{sec:dex}

Regularly a Java program is compiled into a class file
\cite{lindholm2014java}, where each class file holds the code and data
for a single class. Bytecode in class files is stack oriented, while
indices into a constant pool can provide literal values or describe
symbolic references to fields, methods and other classes.

Android introduced the notion of dex files, that hold more than one
class, and are executed by the Dalvik or ART runtimes \cite{art14,
  bornstein2008dalvik}. Various tables in the dex file take the place
of the constant pool and classes may share table entries to reduce
size. Tables may be indexed by dex instructions or from other tables.
For the string table the index may be 32-bit but in general the index
is limited to 16-bits. If there is no room in a table then the
compiler will generate multiple dex files following a convention that
indicates to the runtime that the dex files should be considered one
unit. Classes may not be split over dex files and each dex file is
required to verify separately. It is possible to organize the tables
to either achieve a smaller file, or to optimize start-up speed.

Dex instructions are 2-byte aligned and may be up to 10 bytes in
length \cite{dex_bytecode2020}. They consist of a 1-byte opcode and
then multiple bytes encoding registers used and defined, constants,
branch offsets and symbol table indices. Registers are encoded into
either 4, 8 or 16 bits depending on the instruction. A consequence of
the register encoding is that all instructions can access the low 0
through 15 registers, a subset the higher 16 through 255 registers and
an even smaller set registers 256 through 65,535. To work around these
limitations move instructions can copy a register's value from a high
to a low numbered register or vice-versa.

Instructions that require a variable number of inputs, most commonly
`invokes', have two forms. The first form is to have a list of up to 5
registers, numbered 0 through 15. The second `range' form takes a
base register, any of the 65,535 registers, and a length. An
additional instruction may appear after these instructions to place
the result into a register.

There are smaller two-address forms of some binary instructions, where
the first source register is also the destination.
Table~\ref{tab:adds} shows the four possible encodings of the add
instruction, which has two-address and add-immediate forms. The
two-address form has opcode $B0$ and registers must be in the range 0
to 15. If the add literal is a signed 8-bit value then registers 0 to
255 can be encoded with opcode $D8$, whilst 16-bit literals are possible
with opcode $D0$ but with a limitation that only registers 0 to 15 can
be used. If a register larger than 255 is needed then a move will be
necessary, similarly a constant greater than 16-bit will need generating
into a register.

\begin{table}[hbtp]
  \begin{center}
  \begin{tabular}{|l|l|l|l|}
    \hline
    Opcode & Byte 1    & Byte 2       & Byte 3          \\
    \hline
    90     & vA        & vB           & vC              \\
    B0     & vA and vB & \multicolumn{2}{l|}{\textit{Not used}} \\
    D0     & vA and vB & \multicolumn{2}{l|}{Literal 16} \\
    D8     & vA        & vB           & Literal 8\\
    \hline
  \end{tabular}
  \end{center}
  \caption{Dex file add instruction encodings. vA, vB and vC encode a
  register number}
  \label{tab:adds}
\end{table}

64-bit long and double values occupy adjacent pairs of registers, with
the lowest numbered register being encoded. While the encoding permits
the long or double register to be an odd number, runtimes typically
penalize odd numbered allocations and do not allow them to reside in
64-bit machine registers.

The number of instructions within a method is limited to a 32-bit
value, however, the encoding of exceptions limits most method
locations to being 16-bit.  The number of registers required by a
method can vary and is held in its metadata\footnote{As 65,536
  registers can be encoded, and this is generally greater than the
  number of instructions, a na\"ive register allocator could give each
  instruction its own register. However this ignores issues with
  ranges, longs and doubles and that such an allocation would cause
  frequent stack overflow errors}. Incoming parameters to a method
arrive in the highest numbered registers. For example, if a method has
10 registers and 4 registers for parameters, registers 6 to 9 will
hold the parameters at the start of the method. If a parameter is in
register 16 and then used by an instruction such as `instance-of',
that can only encode registers 0 to 15, the parameter will first need
moving into a lower numbered register.

To summarize the challenges of register allocation of dex instructions,
they are:
\begin{itemize}
  \item Multiple encodings exist for instructions, the shorter forms
    may only be usable for certain registers or literal values.
  \item If all registers 0 to 15 are allocated then instructions may
    fail to be encoded. One solution, that leads to suboptimal code
    generation and is used by the dx and d8 compilers
    \cite{bornstein2008dalvik, d8}, is in register assignment to
    reserve a pool of low numbered `temporary' registers that are
    moved into and out-of to support the high numbered register. The
    approach taken in the presented compiler is to `spill' and
    `fill' low numbered registers into high numbered registers with
    move instructions by live range splitting.
  \item If register allocation increases the number of registers then
    parameters are moved. Move instructions may be necessary to copy
    parameters into registers that can encode instructions with them.
  \item Move instructions can create
    redundant copies of a value, and the register allocator can
    reduce future move instructions by reusing values already
    within an encodable register.
 \item Range operations, or long and double pairs of registers that
    are a range of length 2, require registers to be
    consecutive: \begin{itemize}
      \item Generally the number of registers should be limited by the
        number of live values. If fragmentation occurs then extra
        registers and move operations are necessary to create a
        consecutive range of registers.
      \item The number of move instructions to set up the registers
        for a range operation is proportional to the number of
        arguments to the method invocation. Code size can be minimized
        by coalescing and assigning the output of an instruction to
        the register required for the range operation. This may
        increase fragmentation.
    \end{itemize}
\end{itemize}

While some of these challenges are relatively unique to dex files,
there are similarities with constraints that exist in more general
instruction sets such as for vector register files, spilling-to and
filling-from the stack and two-address encoding.

\subsection{The compiler}
\label{sec:appreduce}

The compiler consists of front ends, middle end optimizations over the
intermediate representation (IR) and back ends.

\subsubsection{Front ends}

The compiler front end can read common Java program wire formats,
namely class files and dex files \cite{lindholm2014java,
  bornstein2008dalvik}. It builds an IR that describes symbols,
classes, fields, methods, annotations and exceptions. Similar to tools
like ProGuard \cite{proguard}, a complete model of a Java program is
held in memory for optimization.

The Java bytecode, or dex instructions, are parsed into a control-flow
graph (CFG) and SSA form using an approach similar to
\cite{Kotzmann:2008:DJH:1369396.1370017} where predecessor basic
blocks are always processed first, to allow $\Phi$ instructions to be
inserted at merges. Loop back-edges and exception catch blocks are
handled pessimistically, with $\Phi$ instructions introduced for all
values and then eliminated via simplification and dead code
elimination. A different approach to SSA form construction is
\cite{braun13} that trades creating and optimizing away unnecessary
$\Phi$ instructions at the cost of recursing over the CFG.

\subsubsection{Middle end optimizations}

A range of both interprocedural and intraprocedural optimizations are
performed on the IR. Being SSA based allows for straightforward global
(between basic block) optimizations, such as common sub-expression
elimination, not possible in ProGuard that uses Java bytecode as an
intermediate form. Dataflow optimizations are performed at the whole
program and method level, with a type lattice that models constants,
type and nullness of references as well as integer ranges. At the
method level, type information is held within the instruction and
always conservatively correct unless a fixed point is being
computed. Fixed points are computed efficiently using Bourdoncle's
approach \cite{bourdoncle93}. By holding type information within the
instruction, pattern matching avoids being specialized upon the size
of, for example, an add - i.e. there are no int-add, long-add,
float-add, double-add instructions as the type information is
sufficient to determine the kind of add necessary.

As the IR is being used for Java, runtime exceptions may occur on many
instructions. Dominance is computed on basic blocks within the
compiler.  If dominance is not required, multiple exception throwing
instructions may be within the same basic block as with the factored
control-flow graph \cite{Choi:1999:EPM:316158.316171}. When dominance
is required, blocks are split at instructions that may throw
exceptions with gotos appended afterward. Breaking blocks is known as
unfactoring the CFG. Often null-pointer exceptions are known not to
occur on the `this' pointer, freshly allocated objects and
constants. The type analysis carries this information forward allowing
instructions to determine whether a runtime exception may occur and
avoiding splitting blocks in cases it is known not possible. It is
further possible to avoid to split blocks when they are not within
try-regions, as long as potentially exception throwing instructions
are not being reordered with memory operations. Java source compilers
will implicitly create try-regions for synchronized blocks, to ensure
that objects are unlocked on all control-flow edges, this can cause
more try-regions than just what is present at the source level.

\subsubsection{Back ends}

The compiler has two back ends, one capable of producing Android dex
files and the other Java class files. This paper focuses on dex file
generation. The dex back end must perform various `legalizations',
such as ensuring synchronized methods begin and end with monitor
acquiring and releasing instructions. It must also form the dex file,
or files, symbol tables. To generate instructions the symbol table
layout, the SSA instructions and a register allocation are
required. The register allocation maps from the SSA instruction to its
allocated register except in the case of a folded constant, when no
mapping will exist. To ensure as compact a representation as possible
various peephole passes are also performed, for example,
opportunistically using a smaller `if' instruction rather than a
`switch'.

\section{The register allocator algorithm}
\label{sec:interval-free-ra}

\subsection{Preparing to allocate}
\label{sec:preparing_to_alloc}

Before register allocation is performed the CFG is unfactored to
ensure instruction level dominance, but only within try-regions as
memory operations will not be reordered. The compiler aims to minimize
the number of registers allocated, with more registers being necessary
when more instructions are live. The `shrink live range' pass aims to
move instructions as close to their uses as possible while maintaining
correct semantics. An area where semantics are by default relaxed is
around out-of-memory errors. An out-of-memory error for a $new$ in
Java should be thrown ahead of the constructor's arguments being
evaluated (section 12.5~\cite{Gosling:2014:JLS:2636997}). The compiler
allows $new$ operations to sink next to the constructor call to avoid
holding the uninitialized object live for the duration of argument
evaluation.

During front end parsing and optimization, all constants were
deduplicated and held within the entry block. This simplifies global
value numbering, done as a part of global common sub-expression and
load/store elimination. It is hoped that constants will be folded into
instructions but certain uses require a register, for example, array
indices and method parameters. Dex has special array filling
operations, but often these are not applicable for the operands of the
array stores. Reusing a constant within a register can reduce code
size but the large live ranges of the constants increases register
usage. If more registers are used then larger instruction encodings,
or moves, may be necessary and this removes the code size benefit of
sharing the register. To reduce register usage, the compiler splits
constants ahead of register allocation. Constants required to be in a
register are duplicated ahead of their use, which if they dominate
later uses may be reused. A na\"ive heuristic is currently used to
determine whether to use one or multiple splits, and that is if the
number of non-folded uses is $\geq 3$ the compiler reuses the
register. The constant's live range can be split further during
register allocation, and constants are favored for live range
splitting as they are trivial to rematerialize.

Liveness is computed as a fixed point on a reverse weak topological
sort order of basic blocks \cite{bourdoncle93}.  Live-ins to an
instruction are calculated from the live-outs unioned with the inputs
to the instruction, less the instruction itself if it was
live-out. Treating $\Phi$s similarly would cause all $\Phi$ inputs to
be live on all predecessors. Instead $\Phi$ inputs are inserted into
the live-outs of predecessor blocks before the fixed point calculation
is performed. An alternate approach, that avoids computing a fixed
point, for strict SSA programs with a loop-nesting forest in two
passes is presented in \cite{brandner:inria-00558509}.

Ahead of register allocation, the CFG has critical edges split and
parallel copy blocks inserted in predecessor blocks. The parallel copy
moves are named $\Phi$-moves within the compiler and are coalesced during
register allocation. A different copy operation is a `swap-move',
which is introduced to add a temporary name to solve the `swap
problem', critical edge splitting solving the `lost copy' problem of
coming out of SSA form \cite{Boissinot:2009:ROT:1545006.1545063}.

In the unfactored CFG, it is common for edges to exception catch
blocks to be critical edges. An edge is critical if the source basic
block has $>1$ successor and the target basic block $>1$
predecessor. Catch blocks generally have the first condition as there
is exceptional and regular control-flow. Catch blocks have a first
instruction to gather the thrown exception known as `GetException' in
the IR. Following critical edge splitting many of these blocks
containing the `GetException' instruction exist, and a $\Phi$
instruction in what was the catch block gathers the different
values. When possible, the compiler merges all of the equivalent
`GetException' blocks ahead of register allocation to reduce the IR's
complexity. This does not introduce a critical edge as there are $>1$
predecessor basic blocks but still just a single successor basic
block.

\subsection{The outer loop}
\label{sec:outer_loop}

A lower bound on the number of registers required for the allocation
is the maximum number of live instructions, ignoring constants that
are folded, and counting live long and double instructions twice - dex
files requiring that long and double values are in two adjacent
registers. In the compiler, register allocation fails when either no
register or no suitable register can be found. For example, if an
`int-to-long' instruction cannot locate a pair of consecutive registers,
for the long result, then the inner register allocator loop fails with
a cause and an iterator at the point of failure. The live instructions
are also known at this point.

Other register allocators for the dex instruction set, such as dx and
d8 \cite{bornstein2008dalvik, d8}, allocate registers in a single pass
and then rename registers during register assignment. During
assignment it is determined if the register constraints for the
allocation hold, and if not moves are introduced. Temporary registers
in the low 15 registers, that always satisfy instruction constraints,
are either reserved prior to allocation, or room created by shuffling
the allocated register numbers up. Adding one temporary may invalidate
constraints on another instruction and so moves may be repeatedly
added until constraints are satisfied.

Our compiler performs allocation over blocks in the final code layout
order. The final code layout uses a queue to determine which block to
visit next. Loop headers and blocks within the same try-region are
placed at the head of the queue. Exception successors or code paths
that terminate at a throw, are placed on the end of the queue. By
falling through to successors code size is minimized. Meta-data size
is minimized by keeping try-regions intact. It is considered unlikely
that exceptions will be thrown or caught.

Exception handling code often has a large number of live variables to
describe a failure, and so this colder code can be a point of an
allocation failure. Allocation failures such as this could be avoided
by a basic block traversal order that prioritizes blocks with a large
number of live values.  The basic block traversal order impacts the
efficiency of the algorithm, for example a reverse post-order
traversal was used in early work to reduce liveness holes
\cite{Poletto:1999:LSR:330249.330250}. Using the final layout ordering
was used to best aid the development of heuristics for allocating a
free register, as will be described in section~\ref{sec:heuristics}, and also
for determining best policies to handle allocation failures, as
described in section~\ref{sec:alloc_failure}. We will consider basic
block traversal order further in section~\ref{sec:ra_improvements}
while section~\ref{sec:analysis} shows measured numbers for allocation
failures with this order in Android code. We will consider the
allocation order of other register allocators in
section~\ref{sec:related}. The next section describes how allocation
failures are recovered from.

\subsection{Allocation failure and live range splitting}
\label{sec:alloc_failure}

A failure to allocate at an instruction, or between basic blocks,
within the main register allocation loop, triggers an allocation
failure with multiple remedies. When the number of registers used is
small, increasing the number of registers is preferred. To ensure
constraints can be met when the number of registers goes beyond 15 or
255, moves are introduced before instructions that can only use the
lower numbered registers. Similarly, parameters that have fixed high
register numbers have moves introduced to provide a degree of freedom
in which register they are encoded for the bulk of the method. For
example, a frequent problem is the `this' pointer arriving in a high
numbered register but then needing to be in registers 0 to 15 for a
field access. Moving the parameters at entry allows a single
dominating move.

Invoke instructions can either have up to 5 arguments in the registers
0 to 15, where longs and doubles require two registers, or take a
range of registers. If there are more than 15 registers, or a method
takes more than 5 arguments, a block of moves is placed to set up the
method invocation. At register allocation time the block is identified
by the type of move, and a set of registers between 0 and 15 or a
contiguous range of registers can be allocated at once. The chosen
method minimizes register moves and may resort to pre-allocation, see
section \ref{sec:pre_allocation}, to achieve this.

Heuristics are used to reduce the chance of live range splitting being
necessary, see section~\ref{sec:heuristics}. However, when it is
necessary the outer loop has provided the failure recovery code with
the point of the allocation failure and the set of live
values. Typically a split is necessary to free up a low numbered
register. A register is freed using a special `spill-move' that is
pre-allocated to a high numbered register, see
section~\ref{sec:pre_allocation}. `Fill-moves' may be necessary to
move the high numbered register into a low numbered register to
satisfy the constraints of an instruction. The split instruction is
selected from those that are live so that the cost, in terms of
inserted moves with some consideration of coalescing, is
minimized. Two possible filling strategies are considered for the
split, introducing a `fill-move' prior to every use that requires it
or having a single `fill-move' that dominates all uses.

\subsection{Register allocation}

Algorithm~\ref{alg:linearscan} shows Poletto and Sarkar's original
linear scan algorithm simplified to remove spilling\footnote{Spilling
  is removed due to the large number of dex registers. To consider the
  new algorithm for a limited size register file and stack, the low
  numbered registers can map to the register file while high numbered
  registers can be considered on the stack.}
\cite{Poletto:1999:LSR:330249.330250}. When spills are necessary
because of constraints, live range splits are performed as described
in section~\ref{sec:alloc_failure}.

\begin{algorithm}[hbp]
\SetKwFunction{ExpireIntervals}{ExpireIntervals}
\SetKwData{Active}{active}
\SetKwData{varI}{i}
\SetKwData{varJ}{j}
\SetKwProg{Fn}{Function}{}{}
\Fn{LinearScanRegisterAllocation}{
  \Active$\leftarrow \emptyset$\;
  \ForEach{live interval \varI, in order of increasing start point}{
    \ExpireIntervals{\varI}\;
    $register[\varI]\leftarrow$ a register removed from pool of free registers\;
    add \varI to \Active, sorted by increasing end point
  }
}
\Fn{ExpireIntervals}{
  \ForEach{interval \varJ in \Active, in order of increasing end point}{
    \If{$endpoint[\varJ] \geq startpoint[\varI]$}{
      return\;
    }
    remove \varJ from \Active\;
    add $register[\varJ]$ to pool of free registers\;
  }
}
 \caption{Linear scan algorithm simplified to not include spilling}
 \label{alg:linearscan}
\end{algorithm}

To summarize the algorithm, it moves forward over intervals in the
order of their start points. The algorithm first expires all intervals
that end before this start point, when an interval expires it is
removed from the set of active intervals and its associated register
marked as free. A value that's not live-out but is live-in has
expired. Once any registers have been freed, a register is selected to
associate with the interval and the interval is made active.

Algorithm~\ref{alg:nointervals} shows the main loop of the new
register allocation algorithm. Some terms in the algorithm are:

\begin{algorithm}[tbp]
  \SetKwFunction{AllocateRegister}{AllocateRegister}
  \SetKwFunction{ExpireIntervals}{ExpireIntervals}
  \SetKwFunction{ExpireIntervalsForInstr}{ExpireIntervalsForInstr}
  \SetKwFunction{StartIntervals}{StartIntervals}
  \SetKwFunction{PreAllocation}{PreAllocation}
  \SetKwData{Active}{active}
  \SetKwData{FutureActive}{future-active}
  \SetKwData{LiveOuts}{live-outs}
  \SetKwData{LiveIns}{live-ins}
  \SetKwData{Uses}{uses}
  \SetKwData{Visited}{visited}
  \SetKwData{varI}{i}
  \SetKwData{varCur}{cur}
  \SetKwProg{Fn}{Function}{}{}
  \Fn{RegisterAllocation}{
    \Active$\leftarrow \emptyset$\;
    \FutureActive$\leftarrow \PreAllocation()$\;
    \Visited$\leftarrow \emptyset$\;
    \LiveOuts$\leftarrow \emptyset$\;
    \ForEach{\varCur is the current basic block from the CFG iterator}{
     \LiveIns$\leftarrow$ set of instructions live into \varCur\;
      \ExpireIntervals{\Active, \FutureActive, \Visited, \LiveOuts, \LiveIns}\;
      \StartIntervals{\Active, \FutureActive, \LiveOuts, \LiveIns}\;
      \ForEach{\varI is the current instruction from forward iteration over \varCur}{
        \If{\varI has inputs}{
          \ExpireIntervalsForInstr{\varCur, \varI, \LiveOuts, \Active, \FutureActive, \Visited, \Uses}\;
        }
        \If{\varI is not a folded constant}{
          \AllocateRegister{\varCur, \varI, \Active, \FutureActive}\;
          \LiveOuts$\leftarrow \varI$\;
        }
      }
      \Visited$\leftarrow \varCur$\;
    }
  }
  \caption{Main register allocation loop}
  \label{alg:nointervals}
\end{algorithm}

\begin{itemize}
  \item{active - a mapping from a register to an instruction that is
    live within it. It is similar to the active interval in
    algorithm~\ref{alg:linearscan}.}
  \item{future-active - a mapping from a register to a set of
    instructions that will occupy it later in the scan. This section
    will consider this set further, and sections
    \ref{sec:pre_allocation} and \ref{sec:ra_improvements} concern its
    use in optimizations.}
  \item{live-ins - the values live into a point in the program
    iteration. The live-ins are computed for basic blocks by a liveness
    pass described in section~\ref{sec:preparing_to_alloc}.}
  \item{live-outs - the values live out of the last basic block or
    instruction. This set is updated as the algorithm moves
    forward. Live-outs are also known for the end of each basic
    block.}
  \item{uses - a mapping from an instruction to a set of instructions
    that use it.}
  \item{CFG iterator - an iterator over the basic blocks of the
    program. Measurements in section~\ref{sec:analysis} are from
    an iterator over blocks in the final code layout order.}
\end{itemize}

The functions within the algorithm will be explained next, but at this
high-level it can be seen that algorithm \ref{alg:nointervals} is
similar to the regular linear scan algorithm \ref{alg:linearscan}.
Rather than iterating over intervals, the algorithm iterates over
basic blocks and the instructions within the basic block. When going
between blocks the live-ins to the block show which instructions need
to be in active. An instruction may be live-out of a basic block but
not live within the block that has been just iterated to. To handle
holes in liveness the algorithm moves an instruction out of the active
set and places it in the future-active set. We term instructions moved
in this way as being paused. It is invariant that an instruction be
absent from both, or in exactly one of, active and future-active.

As with liveness holes, $\Phi$ instructions, and their associated
predecessor block parallel copies, must be placed in either the active
or future-active sets when they are allocated. It is invalid to place
something into the active or future-active set associated with a
register if a \texttt{LiveAtTheSameTime} property is true with an
instruction already within the set. The \texttt{LiveAtTheSameTime}
property is explained below, but in the common trivial case if active
and the future-active are empty then the register can be allocated.

The difference between active and future-active is that active forms
the set of instructions currently occupying registers at the iteration
point in the algorithm. To handle $\Phi$ instructions and liveness
holes, intervals may be merged in a conventional linear scan
algorithm. This algorithm summarizes equivalent information in the
future-active set. To merge two intervals in a conventional linear
scan algorithm, the intervals must not overlap. To place an
instruction in active or future-active, in the presented approach, the
\texttt{LiveAtTheSameTime} property must not hold between the
instructions within the active and future-active sets and the
instruction being added. Characteristics of the programs being
compiled will determine how often \texttt{LiveAtTheSameTime} is
computed, for example, a program consisting of a single basic block
has no $\Phi$ instructions or liveness holes by definition, and
therefore need not use \texttt{LiveAtTheSameTime}. Characteristics of
Android programs and their IR are measured section~\ref{sec:analysis}.

Algorithms~\ref{alg:expire_start_bb} and
\ref{alg:start_and_expire_instr} show the implementation of the
functions used in algorithm~\ref{alg:nointervals}. For the simplicity
of presentation, allocation of $\Phi$, blocks of moves and the
functions \texttt{FreeRegister} and \texttt{PauseRegister} (moving a
register from active to future-active) are not shown. The cost
function used to select the best register is described in
section~\ref{sec:heuristics} and it is also responsible for coalescing
moves.

\begin{algorithm}[hbt!]
  \SetKwFunction{AllocateRegister}{AllocateRegister}
  \SetKwFunction{ExpireIntervals}{ExpireIntervals}
  \SetKwFunction{PauseRegister}{PauseRegister}
  \SetKwFunction{FreeRegister}{FreeRegister}
  \SetKwFunction{StartIntervals}{StartIntervals}
  \SetKwFunction{LiveAtTheSameTime}{LiveAtTheSameTime}
  \SetKwData{Active}{active}
  \SetKwData{FutureActive}{future-active}
  \SetKwData{LiveOuts}{live-outs}
  \SetKwData{LiveIns}{live-ins}
  \SetKwData{Uses}{uses}
  \SetKwData{Visited}{visited}
  \SetKwData{varI}{i}
  \SetKwData{varCur}{cur}
  \SetKwData{varBlock}{block}
  \SetKwProg{Fn}{Function}{}{}
  \Fn{ExpireIntervals}{
    \KwIn{\Active, \FutureActive, \Visited, \LiveOuts, \LiveIns}
    \ForEach{instruction \varI in \LiveOuts but not in \LiveIns}{
      \ForEach{\varBlock in CFG not in \Visited}{
        \If{\varI is live-in}{
          \PauseRegister(\varI, \Active, \FutureActive)\;
          continue outer loop\;
        }
      }
      \FreeRegister(\varI, \Active)\;
    }
  }
  \Fn{StartIntervals}{
    \KwIn{\varCur,\Active, \FutureActive, \Visited, \LiveOuts, \LiveIns}
    \ForEach{instruction \varI in \LiveIns but not in \LiveOuts}{
       \AllocateRegister(\varCur, \varI, \Active, \FutureActive)\;
    }
  }
  \caption{Expire and start intervals between basic blocks}
  \label{alg:expire_start_bb}
\end{algorithm}

\begin{algorithm}[t!]
  \SetKwFunction{ExpireIntervalsForInstr}{ExpireIntervalsForInstr}
  \SetKwFunction{PauseRegister}{PauseRegister}
  \SetKwFunction{FreeRegister}{FreeRegister}
  \SetKwFunction{AllocateRegister}{AllocateRegister}
  \SetKwFunction{LiveAtTheSameTime}{LiveAtTheSameTime}
  \SetKwComment{tcc}{/*}{*/}
  \SetKwData{Active}{active}
  \SetKwData{FutureActive}{future-active}
  \SetKwData{LiveOuts}{live-outs}
  \SetKwData{LiveIns}{live-ins}
  \SetKwData{Uses}{uses}
  \SetKwData{Visited}{visited}
  \SetKwData{varCur}{cur}
  \SetKwData{varI}{i}
  \SetKwData{varJ}{j}
  \SetKwProg{Fn}{Function}{}{}
  \Fn{ExpireIntervalsForInstr}{
    \KwIn{\varCur, \varI, \LiveOuts, \Active, \FutureActive, \Visited, \Uses}
    \ForEach{input value \varJ of instruction \varI}{
      \If{\varJ is folded constant}{
        remove \varJ from \LiveOuts\;
        continue\;
      }
      \If{\varJ not in \LiveOuts}{
        \tcc{\varJ is a duplicate input}
        continue\;
      }
      \If{\varJ is in live-outs of $cur$}{
        \If{basic block of \varJ is not $cur$ and \varJ has no later
          uses within $cur$ than \varI and \varJ is defined after \varI}{
          \tcc{Liveness hole within $cur$}
          \PauseRegister(\varJ, \Active, \FutureActive)\;
          remove \varJ from \LiveOuts\;
        }
        continue\;
      }
      remove \varJ from \LiveOuts\;
      \uIf{\varJ has uses in $cur$ after \varI or \varJ is in the live-ins of a block not in \Visited}{
        \PauseRegister(\varJ, \Active, \FutureActive)\;
      }
      \Else{
        \FreeRegister(\varI, \Active)\;
      }
    }
  }
  \Fn{AllocateRegister}{
    \KwIn{$cur$, \varI, \Active, \FutureActive}
    \uIf{\FutureActive contains \varI} {
      \Active$\leftarrow i$\;
      Remove \varI from \FutureActive\;
    }
    \Else{
      Select lowest-cost register to allocate to \varI ensuring
      \LiveAtTheSameTime is false. If no register is found then
      fail allocation.
    }
  }
  \caption{Expire an interval at an instruction and to allocate a register}
  \label{alg:start_and_expire_instr}
\end{algorithm}

In algorithm~\ref{alg:start_and_expire_instr} the condition on line 9
can only be true for $\Phi$-moves. The condition on line 14 has some
subtlety, it is not sufficient to say there is not a use in an
unvisited basic block as an instruction may be live over a basic
block, but neither used or defined within it. To see whether checking
$cur$ and unvisited blocks is necessary the uses of $j$ are checked to
see whether there is just a use by $i$, whether there are multiple
uses within $cur$ and whether there are uses in unvisited blocks other
than $cur$. Scanning backward to find other future uses is similar to
bottom-up local register allocation \cite{Torczon2012}, but here the
algorithm is performing global rather than local allocation. In
section~\ref{sec:ra_improvements} we will show how the backward scan
can be avoided.

\begin{algorithm}[tbp]
  \SetKwFunction{LiveAtTheSameTime}{LiveAtTheSameTime}
  \SetKwFunction{LiveAtTheSameTimeSameBlock}{LiveAtTheSameTimeSameBlock}
  \SetKwFunction{LiveAtTheSameTimeInBlock}{LiveAtTheSameTimeInBlock}
  \SetKwData{varI}{i}
  \SetKwData{varLhs}{lhs}
  \SetKwData{varRhs}{rhs}
  \SetKwData{varBlock}{block}
  \SetKwData{varFirst}{first}
  \SetKwData{varLast}{last}
  \SetKwData{lhsLiveOut}{lhsLiveOut}
  \SetKwData{rhsLiveOut}{rhsLiveOut}
  \SetKwData{lhsBlock}{lhsBlock}
  \SetKwProg{Fn}{Function}{}{}
  \Fn{LiveAtTheSameTime}{
    \KwIn{\varLhs, \varRhs}
    \If{basic block of \varLhs is the same as \varRhs} {
      return LiveAtTheSameTimeSameBlock(\varLhs,\varRhs,basic block of \varLhs)\;
    }
    return LiveAtTheSameTimeInBlock(\varLhs,\varRhs,basic block of \varLhs) $\lor$ LiveAtTheSameTimeInBlock(\varRhs,\varLhs,basic block of \varRhs)\;
  }
  \Fn{LiveAtTheSameTimeSameBlock}{
    \KwIn{\varLhs, \varRhs, \varBlock}
    $\lhsLiveOut \leftarrow$ is \varLhs in live-outs of \varBlock\;
    $\rhsLiveOut \leftarrow$ is \varRhs in live-outs of \varBlock\;
    \uIf{$\lhsLiveOut \land \rhsLiveOut$} {
      return true\;
    }
    \uElseIf{$\lnot \lhsLiveOut \land \lnot \rhsLiveOut$} {
      $(\varFirst, \varLast) \leftarrow$ search forward in \varBlock until
      encountering \varLhs or \varRhs, \varFirst is the encountered
      instruction and \varLast the other\;
       \ForEach{\varI in backward iteration over \varBlock}{
         \If{$\varI = \varLast$}{return false\;}
         \If{\varFirst is input to \varI}{return true\;}
       }
    }
    \Else{
      \If{\lhsLiveOut}{swap \varLhs and \varRhs\;}
      \ForEach{\varI in backward iteration over \varBlock}{
        \If{$\varI = \varRhs$}{return false\;}
        \If{\varLhs is input to \varI}{return true\;}
      }
    }
  }
  \Fn{LiveAtTheSameTimeInBlock}{
    \KwIn{\varLhs, \varRhs, \lhsBlock}
    \uIf{\varRhs in live-in of \lhsBlock\;}{
      \If{\varRhs in live-out of \lhsBlock\;}{
        return true\;
      }
      \ForEach{\varI in backward iteration over \lhsBlock}{
        \If{$\varI = \varLhs$}{return false\;}
        \If{\varRhs is input to \varI}{return true\;}
      }
    }
    \Else{return false;}
  }
  \caption{Live at the same time}
  \label{alg:live_at_same_time}
\end{algorithm}

Algorithm~\ref{alg:live_at_same_time} shows how
\texttt{LiveAtTheSameTime} is calculated. The algorithm's complexity
is $O($number of instructions within basic block$)$. If the order of
instructions within a basic block is known, then this can be reduced
to $O($number of uses of an instruction$)$ as shown in
algorithm~\ref{alg:live_at_same_time_ordered}. The ordered variant of
the algorithm is used for the measurements in section
\ref{sec:analysis}. The ordering requirement is different from
linearly numbering all instructions as in a conventional linear scan,
just the order within a basic block need be known.

\begin{algorithm}[t!]
  \SetKwFunction{LiveAtTheSameTime}{LiveAtTheSameTime}
  \SetKwFunction{LiveAtTheSameTimeSameBlock}{LiveAtTheSameTimeSameBlock}
  \SetKwFunction{LiveAtTheSameTimeInBlock}{LiveAtTheSameTimeInBlock}
  \SetKwData{Uses}{uses}
  \SetKwData{varI}{i}
  \SetKwData{varLhs}{lhs}
  \SetKwData{varRhs}{rhs}
  \SetKwData{varBlock}{block}
  \SetKwData{lhsLiveOut}{lhsLiveOut}
  \SetKwData{rhsLiveOut}{rhsLiveOut}
  \SetKwData{varFirst}{first}
  \SetKwData{varLast}{last}
  \SetKwData{lhsBlock}{lhsBlock}
  \SetKwProg{Fn}{Function}{}{}
  \Fn{LiveAtTheSameTimeSameBlock}{
    \KwIn{\varLhs, \varRhs, \varBlock, \Uses}
    $\lhsLiveOut \leftarrow$ is \varLhs in live-outs of \varBlock\;
    $\rhsLiveOut \leftarrow$ is \varRhs in live-outs of \varBlock\;
    \uIf{$lhsLiveOut \land rhsLiveOut$} {
      return true\;
    }
    \uElseIf{$\lnot lhsLiveOut \land \lnot rhsLiveOut$} {
      \uIf{\varLhs is before \varRhs}{$(\varFirst, \varLast) \leftarrow (\varLhs, \varRhs)$}
      \Else{$(\varFirst, \varLast) \leftarrow (\varRhs, \varLhs)$}
      \ForEach{\varI in \Uses of \varFirst}{
        \If{$\varBlock = $basic block of $\varI \land \varI$ is after $\varLast$}{
          return true\;
        }
      }
    }
    \Else{
      \If{\lhsLiveOut}{swap \varLhs and \varRhs\;}
      \ForEach{\varI in \Uses of \varLhs}{
        \If{$\varBlock = $basic block of $\varI \land \varI$ is after \varRhs}{
          return true\;
        }
      }
    }
    return false\;
  }
  \Fn{LiveAtTheSameTimeInBlock}{
    \KwIn{\varLhs, \varRhs, \lhsBlock, \Uses}
    \uIf{\varRhs in live-in of \lhsBlock\;}{
      \If{\varRhs in live-out of \lhsBlock\;}{
        return true\;
      }
      \ForEach{\varI in \Uses of \varRhs}{
        \If{$\lhsBlock = $basic block of $\varI \land \varI$ is after \varLhs}{
          return true\;
        }
      }
    }
    return false\;
  }
  \caption{Live at the same time using ordering}
  \label{alg:live_at_same_time_ordered}
\end{algorithm}

\subsection{Pre-allocation}
\label{sec:pre_allocation}

As the target of the compiler is the dex file format, fixed registers
are limited to just the parameters. If the compiler were targeting an
architecture, for example, that required operands for divide
instructions to be in certain registers, then these would be
pre-allocated. Pre-allocation in the compiler means taking certain
instructions and placing them into future-active before the main loop
is ran. As described in section~\ref{sec:alloc_failure}, spills are
pre-allocated as well as block move operations. By allocating these
moves early \texttt{AllocateRegister} can look at the future uses by
moves of a value, and then choose the same register so that the move
does not require an instruction to be generated. This is similar to
fixed intervals in conventional linear scan algorithms
\cite{Wimmer:2005:OIS:1064979.1064998}.

Code size is important for the compiler, and with poor register
allocation move operations can make a substantial contribution. In
pre-allocation of block moves the compiler takes into consideration
that other block moves may duplicate a value in a different
register. The compiler keeps a map from the input instruction, with
$\Phi$ and moves removed, to the allocated register. When choosing
registers to pre-allocate block moves into, knowing that a register
holds a value, or a copy of it, lowers its cost to the pre-allocation
code. The pre-allocation selects registers for block moves that
produce the fewest instructions, or when this is equal, that occupy the
highest numbered registers.

Spills are pre-allocated after all other instructions, as with block
moves, the compiler looks to see if a register will hold a value from
a spill or a spill hold an already pre-allocated value. The register
chosen for a spill is the highest numbered register that minimizes the
number of generated move instructions.

\subsection{Coalescing}
\label{sec:coalescing}

The compiler coalesces by looking backward. When allocating a move,
the top non-$\Phi$ or move input is computed\footnotemark. Iterating over the active
registers the top non-$\Phi$ or move is computed for each active
value. If the same value is seen then that register is considered to
have a lower allocation cost.

\footnotetext{Spill and swap moves are considered somewhat
  differently. If spills are elided we may end up with live values in
  low numbered registers, removing their utility. Swap moves always
  terminate the input search as coalescing a swap move would remove
  the temporary copy of its input.}
  
The compiler also coalesces looking
forward. Section~\ref{sec:pre_allocation} described how block move
pre-allocation selects registers to minimize moves. The compiler also
looks at `future-active', when allocating a register, to see if a move
will later occupy that register that takes this instruction as an
input.

During code generation the top non-$\Phi$ or move input is remembered for
each register. If a move is attempted to be generated to copy the same
value into the register, it is elided. At basic block boundaries, that
are not trivial fall-through cases, the map to elide moves is cleared.

\subsection{Allocation heuristics}
\label{sec:heuristics}

If the allocation of a register is not coalesced the register allocated
is chosen on the basis of cost.

\begin{itemize}
\item{if the instruction being allocated satisfies the constraints for
  a 2-address instruction, inputs in low numbered registers and a
  suitable output register is available, then this lowers the cost of
  using one of the inputs as the output.}
\item{last active description - if debugging is requested then it is
  useful to keep local variables and parameters in the same register
  across allocation, to avoid modifying the debugging metadata. If the
  instruction being allocated's debug information matches the last
  instruction to inhabit a register then the register is considered to
  have a lower cost.}
\item{last active - this set records which instruction previously
  inhabited the active register during the scan. The allocator prefers
  to clobber registers that do not have debug descriptions and those
  bearing object references. We wish to clobber object references as
  the compiler generates more efficient code than previous compilers,
  leading to object references frequently being left in a register and
  in the runtime's interpreter having an extended live
  range. Clobbering the values first removes some of this
  problem\footnote{The compiler's front end introduces instructions to
    track explicit nulls being stored in local variables. This
    information is used to ensure such locals are clobbered by the
    register allocator to avoid an extended lifetime in the runtime's
    interpreter.}.}
\item{the compiler aligns long and double values on even register
  numbers. The original ART ahead-of-time compiler had a simplistic
  mapping of dex registers to fixed ARM registers \cite{art14}. If
  longs or doubles were not in even numbered registers then they would
  be held in the stack frame and loaded and stored for each
  operation.}
\item{rather than allocating high numbered registers to free up low
  numbered registers, the compiler attempts to always allocate regular
  instructions in low numbered registers. Pre-allocation does the
  opposite when selecting registers. Allocation generally produces the
  smallest code size when it uses low numbered registers as these give
  the greatest freedom with instruction encoding.}
\end{itemize}

\section{Analysis}
\label{sec:analysis}

We instrumented the compiler and compiled the Java classes to dex
files for the Android platform from Android Open Source Project
(branch android-9.0.0\_r9) for the full\_x86-user build
\cite{aosp2018}. Measurements were made per method compiled and then
aggregated. The number of methods compiled was 693,166, with 4.851
basic blocks per method on average. The average number of instructions
within a basic block was 7.482, and the average number of registers
allocated per method was 3.532.

\begin{figure}[t!]
  \includegraphics[width=3.6in]{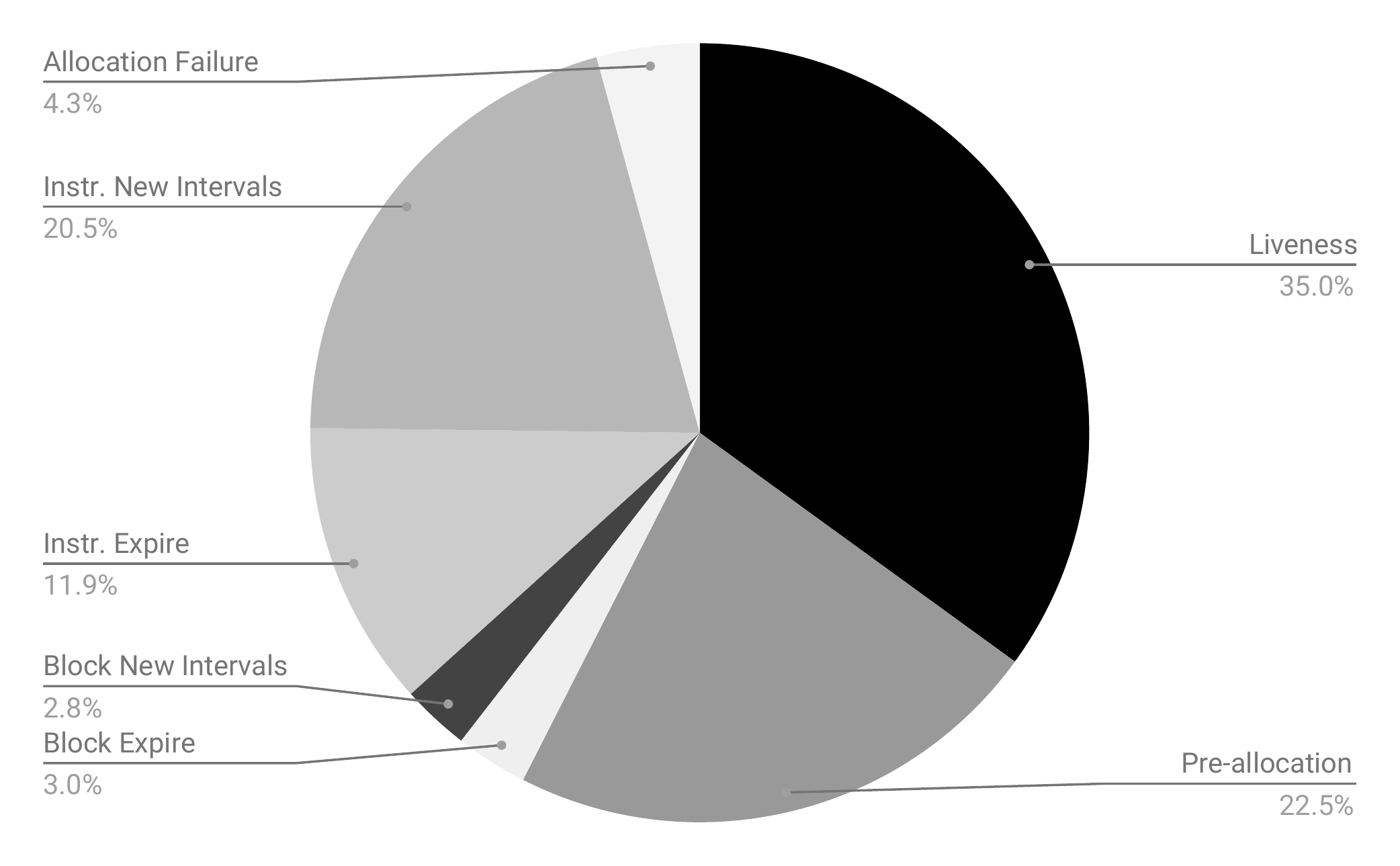}
  \caption{Break down of register allocation time}
  \label{fig:ra-time}
\end{figure}

Fig.~\ref{fig:ra-time} gives a timing breakdown of the parts of the
register allocator\footnote{As individual parts of the register
  allocation execute too quickly to measure with operating system time
  calls, the Intel rdtsc instruction was used and scaled appropriately
  to give wall clock time. Results are the average of 30 runs of
  building Android single threaded on an Intel Xeon E5-2690 at 2.9GHz
  with 64GB of RAM.}.  Almost half the register allocation time is
spent in liveness analysis and pre-allocating instructions before the
main loop is entered. As described in section~\ref{sec:coalescing}
pre-allocation is used to improve coalescing and tries to align blocks
of moves used to set up invoke instructions. This form of
pre-allocation is specific to dex code generation and so such a
proportion of time being spent in pre-allocation need not be necessary
for other targets.  Expiring and then starting intervals using live
sets between basic blocks accounts for 5.806\% of compilation time,
whilst doing the same for instructions accounts for 32.424\% of
register allocation time.

\begin{figure}[b!]
  \includegraphics{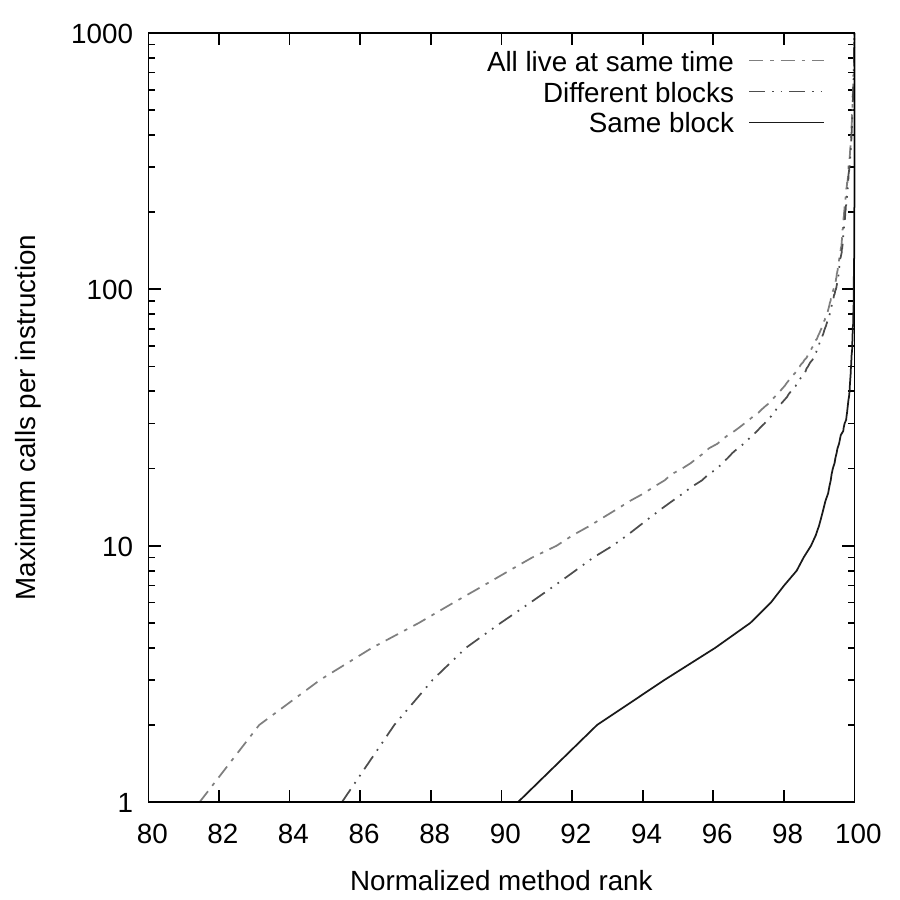}
  \caption{Maximum number of \texttt{AreLiveAtTheSameTime} calls per
    instruction against normalized method rank}
  \label{fig:live_at_same_time}
\end{figure}

\texttt{AreLiveAtTheSameTime} replaces intervals within the register
allocator, of the methods compiled 80.206\% required no calls to this
function. We rank the methods by the number of calls to
\texttt{AreLiveAtTheSameTime} and then plot their maximum calls per
instruction to \texttt{AreLiveAtTheSameTime} broken apart into total,
within the same block or different blocks. The number of instructions
includes retry attempts, and \texttt{AreLiveAtTheSameTime} is counted
for pre-allocation and the main allocation. Calls per instruction is
used as longer methods are expected to make more
calls. Fig.~\ref{fig:live_at_same_time} shows this data. For 90\%,
99\% and 99.9\% of methods, the maximum calls to
\texttt{AreLiveAtTheSameTime} per instruction is less than 10, 70 and
400 respectively. The majority of the calls are to determine liveness
between instructions in different basic blocks. Whilst the large
numbers of calls are for a small fraction of methods, they are
disappointingly large. We found that the outlying cases, with large
numbers of calls, solely comprised of large class initializers that
required a large number of constants. These methods would also
frequently use invokes and range based method invocation that made use
of pre-allocation. Section~\ref{sec:preparing_to_alloc} described the
live range splitting for constants as being na\"ive and more splitting
would reduce calls to \texttt{AreLiveAtTheSameTime} but possibly at
that the cost of code quality. Section~\ref{sec:ra_improvements}
describes how the number of calls could be reduced using different
block and instruction iteration strategies.

\begin{figure}[t!]
  \includegraphics[width=3.4in]{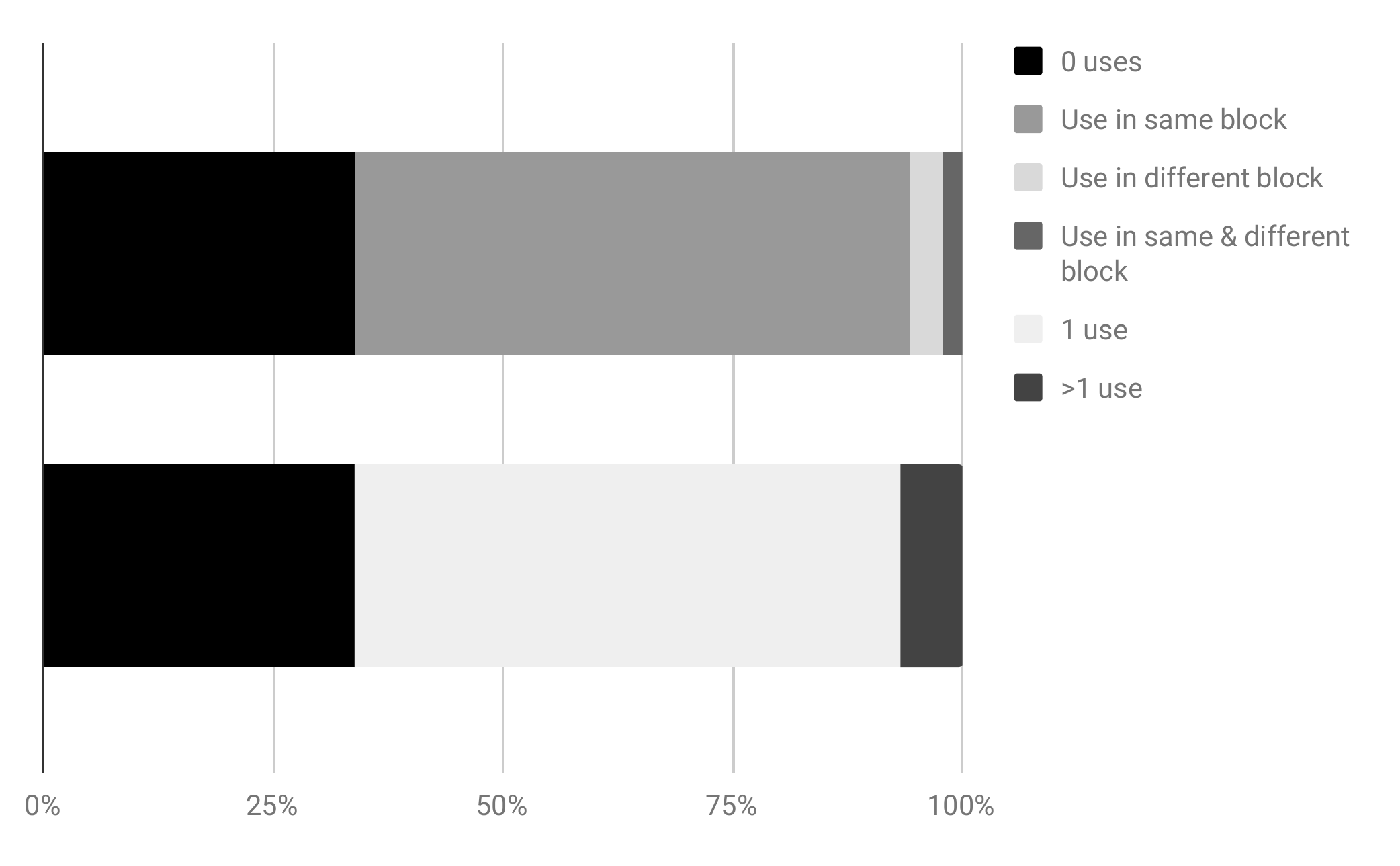}
  \caption{Kinds of instruction uses}
  \label{fig:uses}
\end{figure}

On average each instruction is used by less than one other
instruction, 0.793 for our test data. Fig.~\ref{fig:uses} shows the
breakdown of what kind of instruction uses exist. 33.819\% of
instructions have no uses, 60.452\% are used in just their defining
basic block and 59.362\% have just one use. We conclude that over 90\%
of intervals for instructions are trivial - a single use, or uses just
within the defining basic block.


\begin{figure}[hbt!]
  \includegraphics[width=3.2in]{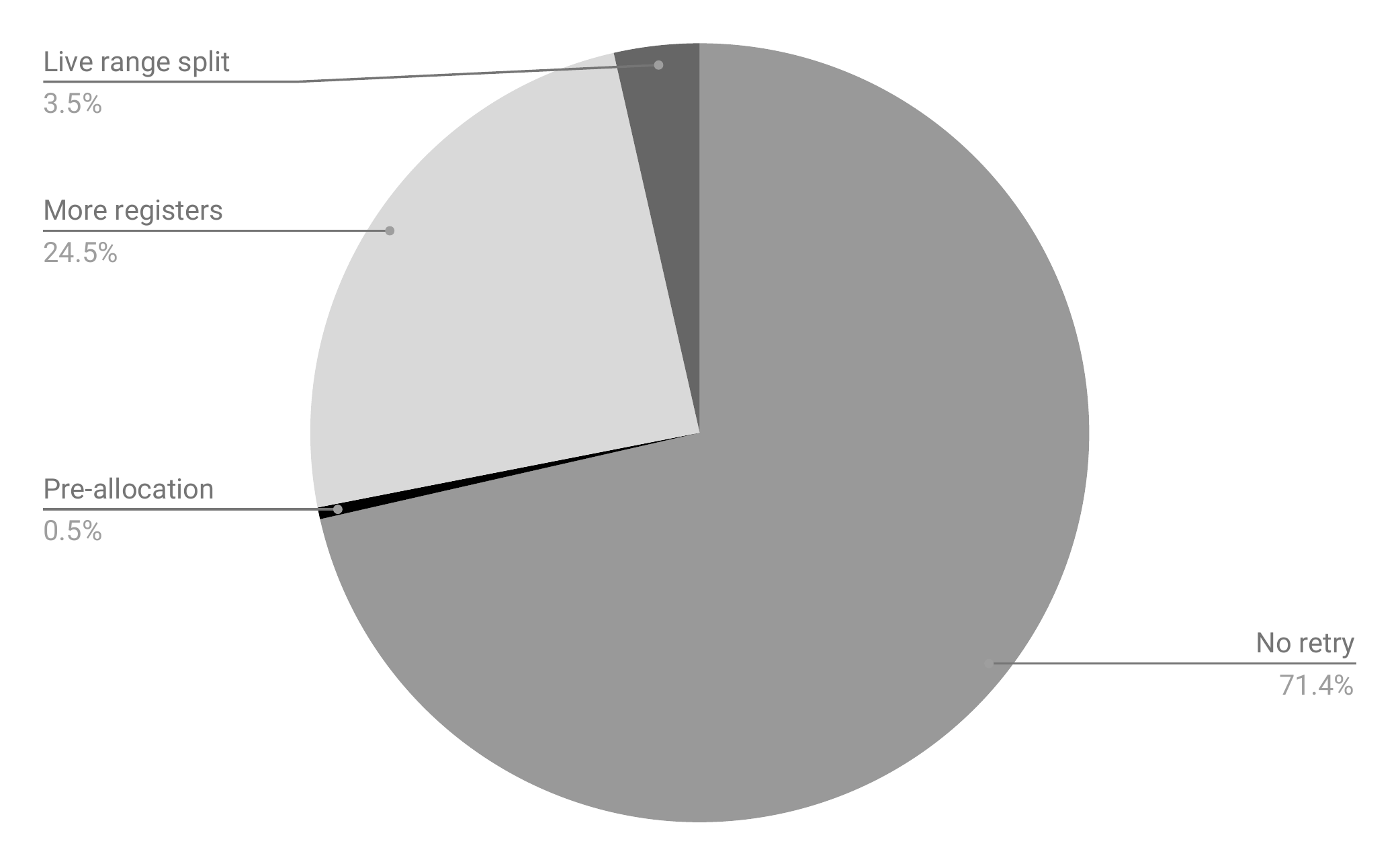}
  \caption{Kinds of retry attempt for the outer allocation loop}
  \label{fig:retries}
\end{figure}

The outer loop will retry allocation for pre-allocation to aid
coalescing, if there are insufficient registers\footnote{As the
  compiler does not perform a register assignment pass, to minimize
  moves, it needs the exact the register numbers to compute registers
  for parameters.} and to split live ranges to free up low numbered
registers. Fig.~\ref{fig:retries} gives a breakdown of the kinds of
retry attempt that cause the outer loop to restart. In 81.330\% of the
methods compiled no retrying was necessary, and for a further 18.195\%
allocation was only retried as there were insufficient registers to
meet instruction constraints\footnote{These numbers differ from
  figure~\ref{fig:retries} as a method may retry allocation more than
  once.}. Pre-allocation is 0.5\% of all retry attempts, and is used
to introduce block moves. Initially block moves are considered not to be
necessary for operations with a small number of arguments. If
insufficient low numbered registers are available then the block moves
are pre-allocated in higher numbered registers so that they are
visible to forward coalescing and allocation retried.


\section{Improvements}
\label{sec:ra_improvements}

By using the final code layout order to allocate registers, it was
hoped that the best decisions for register allocation and how to
handle allocation failures could be made. For example, in the case of
trivial fall-through between blocks, information to guide clobbering
heuristics could be retained. However, final code layout order is not
optimized to avoid liveness holes. For example, a value could be used
in a catch block at the end of the iteration order, and consequently
the future-active register can end up holding the value during most of
the allocation. All allocations to that register will then require a
\texttt{LiveAtTheSameTime} test. Using a reverse post-order traversal
would likely reduce this problem, as could considering loops and loop
nesting \cite{bourdoncle93, brandner:inria-00558509}.

A different problem is in establishing whether the use of a value is
the last use. The expire functions must either iterate over unvisited
blocks, or scan uses within a block. Hecht and Ullman established that
CFG traversal order can be informed by dominance
\cite{Hecht:1973:ASA:512927.512946}. Going in a dominance order would
still not simplify the last use question, but reversing the order and
iterating backward through instructions would. In a such a traversal,
a use would allocate a register whereas arriving at the definition
would indicate that the register is now free, all other live range
holes would indicate a pause.

A backward traversal would also allow block moves to be allocated
before their inputs. Pre-allocation of block moves to aid coalescing
causes greater use of the future-active sets, and more
\texttt{LiveAtTheSameTime} tests.

Whilst improving runtime performance is possible, it is also possible
to improve the quality of code generation. Other interval based
allocators, such as the greedy register allocator in LLVM
\cite{olesen2011register}, do not allocate registers in order of start
point but instead use priority queues to determine the ordering of
intervals to allocate. A similar approach is applicable to this
compiler where instead of a priority queue of intervals, a priority
queue of instructions can be formed. Allocation would be similar to
pre-allocation, see section~\ref{sec:pre_allocation}. Fitting
instructions into registers in an ad hoc manner may motivate the
construction of intervals to lower the cost of
\texttt{AreLiveAtTheSameTime}, but other heuristics may achieve a
similar effect. If intervals were a performance improvement then they
could be lazily constructed as common IR patterns, such as 33.819\% of
instructions having no uses, would still benefit from the approach
presented here.

\section{Related work}
\label{sec:related}

From the beginning of linear scan register allocation, the removal of
interval creation has been in the mind of developers
\cite{Poletto:1999:LSR:330249.330250}. Poletto and Sarkar's approach
was based on strongly connected components and produced an
approximation of liveness that impacted code quality in large
benchmarks, in the extreme making them 6.8 times slower. They reasoned
that the approach may be suitable when quickly compiling small
functions. Similar to small functions are `trivial traces' for which
Eisl et al. present a bottom-up register allocator that is faster than
linear scan \cite{Eisl:2017:TRA:3132190.3132209}. However, given the
lowering of code quality this allocator is used for the compilation of
traces of lower importance to peak performance. Sarkar and Barik
contributed extended linear scan to improve code generation quality to
being equivalent to graph coloring while retaining linear scan's
performance \cite{Sarkar:2007:ELS:1759937.1759950}.

Treescan register allocation seeks to exploit properties of SSA form
and avoids the creation of intervals \cite{Colombet11}. A primary use
for the Treescan register allocator was envisioned to be in
just-in-time compilers where fast compile times could be beneficial
over code quality. A code quality issue in Treescan is that
information about what will need to occupy a register is not kept,
instead moves are inserted and possibly loop backedges broken so that
fixed and $\Phi$ register requirements can be met (this is known as
repairing). Pre-coalescing looks to reduce this code generation
issue. If an ability to look-ahead is given to Treescan, such as with
`future-active' sets, then the behavior will match the register
allocator here with a reverse post order block traversal
strategy. However, as described in section~\ref{sec:ra_improvements},
for minimal code generation time a post-order traversal iterating
backward through instructions in the basic block may be faster as the
last use need not be considered to expire an interval.

Treescan is built around a fast liveness analysis \cite{Boissinot08}
whereas the approach here uses a global liveness described in
section~\ref{sec:preparing_to_alloc}. In detecting
`live-at-the-same-time' between instructions the approach has
similarities to bottom-up local register allocation
\cite{Torczon2012}. Whilst simplistic this approach keeps the
intermediate representation and side metadata down to a minimum, and
as such has similarities with efforts in compilers such as
sea-of-nodes representations \cite{click95}.

Previous work has contrasted linear scan and graph coloring forms of
register allocation \cite{rong09, Sarkar:2007:ELS:1759937.1759950}. As
the interference graph is synonymous with graph coloring, liveness
intervals have become synonymous with linear scan allocators, and this
work shows how a well performing linear scan allocator can be made
without liveness intervals, with an attempt to better guide decisions
around live range splitting and coalescing.

Optimal register allocation has looked to use cost models to compute a
cheapest possible register allocation \cite{hames06,
  Lozano:2016:RAI:2892208.2892237}. Cost models are related to
heuristics, for example, LLVM's greedy register allocator places a
cost on allocating into callee-save registers as the prologue and
epilogue will need to save and restore them
\cite{olesen2011register}. A cost model can model spilling and filling
the callee-save, and if given enough freedom, consider spilling and
filling in more than just the prologue and epilogue. The compile time
performance of optimal register allocation has meant that it has not
been widely adopted. Learning good heuristics for scan based
allocation, from an optimal allocation, could enable a compromise in
produced code quality and compile time register allocator performance.

The heuristics for coalescing presented in
section~\ref{sec:coalescing} are similar to register hints introduced
by Wimmer and M\"{o}ssenb\"{o}ck \cite{Wimmer:2005:OIS:1064979.1064998,
  Wimmer:2010:LSR:1772954.1772979}, in that they are a cheap heuristic
hoping to reduce code size. Similarly another simple heuristic is that
they avoid fills and spills within loops to improve runtime
performance. In this work code size was more important and so a
similar heuristic was not used in live range splitting.

Treescan aimed to minimize repairing through pre-coalescing and used
an interference like analysis based on SSA form to do this
\cite{Boissinot:2009:ROT:1545006.1545063, Budimlic02}. Efficient
coalescing and SSA register allocation are also tackled by Braun et
al. \cite{Braun10}. Unlike those works, this work does not coalesce
based on representations or analyses. Coalescing decisions are
considered during the allocation pass and splits introduced at the
point the machine constraints are reached. In dex code redundant
copies of values are common, consider an argument to a method that is
passed many times and must appear at different argument positions. The
allocator elides copies reusing existing duplicates. The lightweight
approach presented has allowed for unique machine optimizations to be
applied and it is a significant virtue of the approach, brought about
by the desire to avoid implementation complexity.

\section{Conclusions}
\label{sec:conclusions}

The popularity of linear scan register allocation led to the
popularity of intervals to determine interference between values being
allocated. This paper has shown a scan based register allocation
algorithm without intervals, removing a significant cost from linear
scan allocation while retaining its global register allocation
property. The paper has presented this algorithm within a production
compiler mapping between virtual machines, aiming at minimizing code
size. A compile time breakdown is given showing the new algorithm to
have low cost, as well as features of the IR that justify the choice
of the algorithm - specifically that over 90\% of live ranges are
trivial and over 80\% of methods can be register allocated without
consideration of what will later be in the register. The analysis of
what happens in the 20\% shows that for Java code the costs increase
for a small percentage of methods, in particular class initializers.

The approach is lightweight allowing for novel coalescing
optimizations. The paper has also considered how the algorithm may be
improved in both runtime performance and code generation quality. The
lack of constraints on block order allows for trade-offs but may put
pressure on `future-active' sets and `live-at-the-same-time' calls as
a consequence. However, the allocation algorithm is unique among SSA
register allocators that avoid graphs and intervals in that these
trade-offs can be fully explored.

In the context of dex code generation the approach did not seek to
minimize register allocation time and focused on code quality, for
example by retrying register allocation when machine constraints were
reached. Even with the sub-optimal basic block iteration order and
retrying, the cost of the global register allocation is less than two
times the liveness computation. With low register allocation cost, we
believe the approach to be broadly useful for fast and low memory
overhead compilation such as for just-in-time compilers.

\section{Acknowledgments}

The author wishes to thank Google for their support. This work
wouldn't have been possible without the encouragement, feedback,
support and energy of Lirong He, Andreas Gampe, Jisheng Zhao, Raj
Barik, Vivek Sarkar, Ross McIlroy, Martin Maas, Jason Parachoniak,
Michael Quigley, Nicholas Tobey, Arnaud Venet, Shai Barak, Eddie
Aftandilian, Jeremy Manson, Kevin Bierhoff, Liam Cushon, Chuck
Rasbold, Ivan Posva, Danny Berlin, Sanjay Ghemawat, Jim Stichnoth and
Diego Novillo. The author would also like to thank Matthias Braun and
Lang Hames for enlightening conversations.

\bibliographystyle{ACM-Reference-Format}
\bibliography{paper}

\end{document}